\documentclass[twocolumn,showpacs,preprintnumbers,amsmath,amssymb]{revtex4}

\usepackage{graphicx}
\usepackage{dcolumn}
\usepackage{bm}

\def\beq{\begin{eqnarray}}    
\def\eeq{\end{eqnarray}}      

\begin{document}

\title{Studying the decay of the vacuum energy with the
observed density fluctuation spectrum}

\author{Reuven Opher}
\email{opher@astro.iag.usp.br}
\author{Ana Pelinson}
\email{anapel@astro.iag.usp.br} \affiliation{IAG, Universidade de
S\~{a}o Paulo, Rua do Mat\~{a}o, 1226 \\
Cidade Universit\'aria, CEP 05508-900. S\~{a}o Paulo, S.P.,
Brazil}

\date{\today}

\begin{abstract}
We investigate here models that suggest that the vacuum energy
decays into cold dark matter (CDM) and show that the density
fluctuation spectrum obtained from the cosmic microwave background
(CMB) data together with large galaxy surveys (e.g., the Sloan
Digital Sky Survey), puts strong limits on the rate of decay of
the vacuum energy.
CDM produced by a decaying vacuum energy would dilute the density
fluctuation spectrum, created in the primordial universe and
observed with large galaxy surveys at low redshifts.
Our results indicate that the decay rate of the vacuum energy
into CDM is extremely small.
\end{abstract}

\maketitle

\section{Introduction}

Bronstein (1933) was the first to introduce the idea that the
vacuum energy could decay by the emission of matter or radiation
\cite{brons}. Later, a wide variety of phenomenological models for
the decay of vacuum energy were suggested (e.g.,
\cite{vdec1,vdec2,vdec3,vdec4,free,birk,JLimabirk,overd,peeb,coop,shapiro,JLima}).
A particularly interesting model was that of Freese et al.
\cite{free}, who assumed that the vacuum energy density $\rho_v$
is related to the relativistic matter density $\rho_r$ as
$x\equiv\rho_{v}/(\rho_{r}+\rho_{v})=constant$, where
$\rho_{r}=\rho_{e}+\rho_{\gamma}+\rho_ {\nu}$ ($e^{+}\,e^{-}$
pairs, photons, and $N_{\nu}$ species of neutrinos). They noted
that $x$ must be less than $0.07$ in order to have produced the
observed ratio of baryons to photons in the universe in the
nucleosynthesis epoch.
In their model, the vacuum energy density,
$\rho_v=[x/(1-x)]\rho_r$, decreases at a similar rate as does
$\rho_r$ since $x$ is constant.

Birkel and Sarkar \cite{birk} also studied the model of Freese et
al. However, they assumed that the vacuum energy only decays into
photons: $x\equiv\rho_{v}/(\rho_{\gamma}+\rho_{v})$. They also
assumed that $x$ was constant during the evolution of the
universe. From the condition that the decay of the vacuum energy
density must be consistent with primordial nucleosynthesis
abundances, they found an upper $x$ limit, $x_{\rm max}=0.13$,
which corresponds to $\rho_{v}<4.5\times 10^{-12}\,{\rm GeV}^4$
(for a nucleon-to-photon ratio $\eta\simeq 3.7\times 10^{-10}$).
This value for $\rho_{v}$ is orders of magnitude greater than the
present value obtained from recent Type Ia supernovae data,
$\Lambda_0=\Omega^{\,0}_{\Lambda}\,\rho_c^0\simeq 6\, h_0^2\times
10^{-47}\, {\rm GeV}^4$ \cite{Perm99,Riess98}.
Since $\rho_{\gamma}\ll \rho_v$ at present, $x\sim 1$, which is
much greater than the value $x_{\rm max}=0.13$, found by Birkel
and Sarker and $x_{\rm max}=0.07$, found by Freese et al. Thus,
these constant $x$ models are inconsistent with observational
data.

Nonsingular deflationary cosmology models, considered by Lima and
Trodden \cite{JLimabirk}, were also discussed by Birkel and Sarkar
\cite{birk}. These models are a generalization of the model of
Freese et al., with $x$ given by
$x^{\prime}\equiv\rho_v/(\rho_v+\rho_m+
\rho_r)=\beta+(1-\beta)(H/H_I)$, where $\beta$ is a dimensionless
constant of order unity, $\rho_m$ is the nonrelativistic matter
density, and $H_I$ is the inflationary Hubble parameter.
In this generalized model, $x^{\prime}$ is not strictly a constant
since it varies with $H$. It only becomes constant when $H\ll H_I$
(i.e., for times much greater than the inflation era).
Lima and Trodden \cite{JLimabirk} required that $\beta\geq 0.21$
due to the age of the universe. However, Birkel and Sarkar argued
that these nonsingular deflationary cosmological models are
invalid since $\beta<0.13$ from primordial nucleosynthesis data.

Overduin et al. \cite{overd} studied the $x$ parameter of Freese
et al. using a step function, $x(t)=x_r$\, when $t<t_{eq}$, the
equipartition time when the matter density is equal to the
radiation density, and $x(t)=x_m$ when $t> t_{eq}$. They found
that $x$ can not exceed 0.001, in order not to distort the CMB
spectrum.
Since, at present ($z\sim 0$), $x$ is close to unity, this value
is inconsistent with a constant $x$.

In this article, we do not assume a constant $x$ as do Birkel and Sarker,
Overduin et al. and Lima and Trodden (for times much greater than the
inflation era). We assume only that the vacuum energy decays into CDM as a
function of the redshift between the recombination era and the present. If
the vacuum energy decays into CDM, increasing $\rho$, the
$\delta\rho/\rho$ spectrum observed at low redshifts would have been
diluted and the $\delta\rho/\rho$ would have been bigger at the
recombination era. We examine to what extent the vacuum energy density can
vary with redshift from the recombination era ($z\sim 1070$) to the
present ($z\sim 0$), based on recent data of the CMB anisotropies.

The density fluctuations obtained by the 2dF galaxy redshift survey
(2dFGRS) were compared with the measurements of the CMB anisotropies by
Peacock et al. \cite{2df}. They analyzed the average value of the ratio of
the galaxy to the matter power spectra, defining a bias parameter,
$b^2\equiv{P_{gg}(k)}/{P_{mm}(k)}$, over the range of wave numbers
$0.02<k<0.15\,h\,{\rm Mpc}^{-1}$. The scale-independent bias parameter at
the present epoch, was found to be $=1.10\pm 0.08$. They also found that
the matter power spectrum, derived from the galaxy distribution $P_{gg}$
data, differs from that derived from the CMB data by no more than $10\%$
\cite{2df}. Using this result we examine the decay rate of the vacuum
energy into CDM.

The paper is organized as follows. In $\S$ II, we discuss the
decay of the vacuum energy into CDM. Conclusions are presented in
section III.

\section{Vacuum energy decaying into CDM}

A decaying vacuum energy into CDM increases the density of matter
${\rho }$, diluting the ($\delta\rho/\rho$) spectrum.
Consequently, a larger density fluctuation spectrum
$(\delta\rho/\rho)^2$ is predicted at the recombination era
($z_{\rm rec}=1070$) by the factor

\begin{equation}
F\equiv \left[ \frac{\overline{\rho}_M
\,(z)}{\overline{\rho}_M\,(z)-\Delta \rho (z)}\right]^{2}
\,\Big|_{z=z_{\rm rec}}\, \,, \label{fatorf}
\end{equation}
where
\begin{equation}
\overline{\rho}_M\,(z)=\rho _{c}^{0}\,\left(
1+z\right)^{3}\Omega_{M}^0  \label{rhostand}
\end{equation}
is the matter density for a constant vacuum energy density,
$\rho_c ^0 \equiv 3\,H_{0}^{2} / (8\,\pi \,G) \simeq
1.88\,h_{0}^{2}\times 10^{-29}\, g\,cm^{-3}$ is the critical
density, and $\Omega_{M}^0$ is the normalized matter density,
$\Omega_{M}^0=\rho^0_{M}/\rho^0_{c}$ ($\sim 0.3$).

The difference between the matter density $\bar{\rho}_M$ and the matter
density $\rho_{Mv}$ predicted by the model in which the vacuum energy
decays into matter, is
\begin{equation}
\Delta \rho(z) =\overline{\rho}_M(z)-{\rho}_{Mv}(z)\,.
\label{deltarho1}
\end{equation}

The density $\rho_{Mv}(z)$ is normalized at redshift $z=0$
$\left(\rho_{Mv}(z=0)\equiv{\rho^0}_{M}\right)$. In order to
describe the transfer of the vacuum energy $\rho_{\Lambda}$ into
matter $\rho_{Mv}$ \cite{brons}, we use the conservation of energy
equation,
\begin{equation}
\dot{\rho}_{\Lambda}+\dot{\rho}_{Mv}+3\,H\,(\rho_{Mv}+P_{Mv})=0\,,
\label{mequat}
\end{equation}
where $P_{Mv}$ is the pressure due to $\rho_{Mv}$. For CDM, we
have $P_{Mv}=0$.

There exists an extensive list of phenomenological $\Lambda$-decay laws.
Several models in the literature \cite{waga} are described by a power law
dependence
\begin{equation}
\rho_{\Lambda}(z)= \rho_{\,\Lambda}^{\,0}\,(1+z)^{\,n}\,, \label{mainL}
\end{equation}
where $\rho_{\,\Lambda}^{\,0}\equiv \rho_{\Lambda}(z=0)$, which we
investigate here. Chen and Wu \cite{vdec3}, for example, argued that
$n=-2$ from dimensional considerations and general assumptions in line
with quantum cosmology. In particular, they noted that this time variation
of $\rho_{\Lambda}$ leads to the creation of matter with a present rate
which is comparable to that in the steady-state cosmology.

Following Peebles and Ratra \cite{peeb}, the solution for the matter
density has the form
\begin{equation}
\rho_{Mv}(z)=A\,(1+z)^3+B\,\rho_{\Lambda}(z)\,,
\label{matter}
\end{equation}
where $A$ and $B$ are  unknown constants. Using Eqs.(\ref{matter})
and (\ref{mainL}) in Eq.(\ref{mequat}), the dependence of
$\rho_{Mv}$ as a function of $n$ in Eq.(\ref{mainL}) is
\begin{equation}
\rho_{Mv}(z)=\rho_{Mv}^0 (1+z)^3-\left( \frac{n}{3-n} \right) \,
\rho_{\Lambda}^0\, \left[\,(1+z)^3-(1+z)^n\right]\,.
\,\label{rhopeeb}
\end{equation}
Using Eqs. (\ref{rhostand}) and (\ref{rhopeeb}) in Eq.(\ref{deltarho1}),
we find from Eq.(\ref{fatorf}) that
\begin{equation}
F=\left[{1-\left( \frac{n}{3-n} \right)\,
\left(\frac{\rho_{\Lambda}^0}{\rho_{Mv}^0}\right)\,
\left[\,1-(1+z)^{n-3}\right]}\right]^{-2}\,. \label{facpeeb}
\end{equation}

If, as discussed in section I, the density power spectrum from
observations can be increased by no more than $10\%$ due to the decay of
the vacuum energy, we then have a maximum value for the $F$ factor $F_{\rm
max}=1.1$. This maximum value gives $n_{\rm max}\approx 0.06$.

It is interesting to compare the vacuum energy density in the
primordial nucleosynthesis era $\rho_{\Lambda PN}$, with the above
value of $n_{\rm max}$ for the vacuum energy decay dependence
given by Eq.(\ref{mainL}). In the nucleosynthesis era ($z\sim
10^{10}$), we find that $\rho_{\Lambda
PN}=\rho_{\Lambda}^0\,10^{10\,n}=\rho_{\Lambda}^0\,10^{0.6}$.
Using $\rho^0_{\Lambda}\cong 6\,h_0^2\times 10^{-47}\,{\rm GeV}^4$
\cite{Perm99,Riess98}, we obtain $\rho_{\Lambda PN}\simeq
2\,h_0^2\times 10^{-46}\,{\rm GeV}^4$. This is many orders of
magnitude smaller than the maximum value $\rho_{\Lambda PN}\simeq
4.5\times 10^{-12}\,{\rm GeV}^4$, obtained by Birkel and Sarkar
\cite{birk} or $\rho_{\Lambda PN}=1.1\times 10^{-12}\, {\rm
GeV}^4$, obtained by Freese et al. \cite{free}.

As noted above, Eq.(\ref{mainL}) describes a power law dependence of
$\rho_{\Lambda}$ on the cosmic scale factor $a=(1+z)^{-1}$. Recently,
Shapiro and Sol\`a \cite{shapiro} suggested a first order time derivative
dependence of $\rho_{\Lambda}$ on $a$:
$\rho_{\Lambda}\propto(da/dt)^2/a^2\equiv H^2$, where $H$ is the Hubble
parameter. They were motivated by the renormalization group equation that
may emerge from a quantum field theory formulation. They find a redshift
dependence of the cosmological constants
\begin{equation}
\Lambda(z;\nu)= \Lambda_0+\rho_c^0\,\,f(z,\nu)\,, \label{Lnu}
\end{equation}
where $\Lambda(z=0)=\Lambda_0$, $k=0$, and
\begin{eqnarray}
f(z)=\frac{\nu}{1-\nu}\,\left[\left(1+z\right)^{3(1-\nu)}-1\right]\,.
\label{funcz}
\end{eqnarray}
The dimensionless parameter $\nu$ in Eq.(\ref{funcz}) comes from the
renormalization group
\begin{equation}
\nu\equiv \frac{\sigma}{12\,\pi}\,\frac{M^2}{M_P^2}\,, \label{nu}
\end{equation}
where $\,\sigma M^2$ is the sum of all existing particles (fermions with
$\sigma=-1$ and bosons with $\sigma=+1$). The range of $\nu$ is $\nu \in
(0,1)$ \cite{shz}.

We take Eqs.(\ref{Lnu}) and (\ref{funcz}) as a generic form for studying
the decaying vacuum energy into CDM depending on a single parameter $\nu$,
regardless of its theoretical origin.

Using Eqs.(\ref{Lnu}) and (\ref{funcz}), the matter density can be
obtained as a function of $z$ and $\nu$ in the matter era
($P_{Mv}=0$):
\begin{equation}
\rho_{Mv}(z;\nu) \,=\,\rho_{Mv}^0 \,(1+z)^{3(1-\nu)}\,.
\label{Rnu}
\end{equation}
Using Eqs.(\ref{Rnu}) and (\ref{rhostand}) in
Eq.(\ref{deltarho1}), the matter density difference $\Delta\rho$
at the recombination era is
\begin{equation}
\Delta\rho=\rho _{Mv}^{0}\,\left( 1+z_{\rm rec}\right) ^{3} \left[
\,\left( 1+z_{\rm rec}\right) ^{-3\nu}-1\right]\,. \label{delta}
\end{equation}
The factor $F$ modifying the density power spectrum is obtained,
substituting Eqs.(\ref{rhostand}) and (\ref{delta}) in
Eq.(\ref{fatorf}):
\begin{equation}
F=(1+z_{\rm rec})^{6\nu} \,. \label{runcos}
\end{equation}
Using $z_{\rm rec}\approx 1070$ and the maximum value of $\nu$ allowed in
\cite{shapiro}, $\nu=0.1$, we find that $F\approx 66$. For the canonical
choice $M^2=M_P^2$ in Eq.(\ref{nu}), $\nu\approx 2.6\times10^{-2}$ and we
obtain $F\approx 3$.

As noted above, observational data indicate that $F_{\rm max}=1.1$. From
this, we predict that $\nu_{\rm max}$ has a very small value, $\nu_{\rm
max}\approx 2.3\times 10^{-3}$.


\section{Conclusions}

We showed how the observed CMB and large galaxy survey data limit
the vacuum energy decay rate into CDM between the recombination
era and the present. When the vacuum energy decays into CDM,
$\delta\rho/\rho$ is diluted. The density fluctuation spectrum is
amplified by a factor $F$ at the recombination era. From
observations, the density power spectrum can be amplified by no
more than $10\%$ and the maximum value for $F$ is $F_{\rm
max}=1.1$.


We investigate two forms for the decay of the vacuum energy
$\rho_{\Lambda}$:

\noindent 1) A general dependence on the cosmic scale factor $a$:
$\rho_{\Lambda}(z,n)\propto a^{-n}$; and\\
\noindent 2) A quadratic first derivative time dependence on the cosmic
scale factor $a$: $\rho_{\Lambda}(z,\nu)\propto (da/dt)^2/a^2\equiv H^2$,
where $\rho_{\Lambda}=const$ for the parameter $\nu=0$. We place upper
limits on the values of $n$ and $\nu$.

We find that the decay of the vacuum energy into CDM as a scale factor
power law $\rho_{\Lambda}\propto (1+z)^{n}$, gives a maximum value for the
exponent $n_{\rm max}\approx 0.06$.
Similarly, for a parametrized vacuum decay into CDM model with
$\Lambda(z;\nu)=\Lambda_0+\rho_c^0\,[\nu/(1-\nu)]\,
[\left(1+z\right)^{3(1-\nu)}-1]\, \,$, where $\rho_c^0$ is the present
critical density, we have an upper limit on the $\nu$ parameter, $\nu_{\rm
max}=2.3\times 10^{-3}$.

Extrapolating $\rho_{\Lambda}$ back to the primordial nucleosynthesis era
with a dependence $\rho_{\Lambda PN}\propto (1+z)^{n}$, we examine the
predicted value for the vacuum energy density $\rho_{\Lambda PN}$ for the
maximum value $n_{\rm max}=0.06$. We obtain a maximum value for the vacuum
energy $\rho_{\Lambda PN}\simeq 2\,h_0^2\times 10^{-42}\,{\rm GeV}^4$.
This can be compared with the Freese et al. \cite{free} maximum value,
$\rho_{\Lambda PN}=1.1\times 10^{-12}\, {\rm GeV}^4$, and the Birkel and
Sarkar \cite{birk} maximum value, $\rho_{\Lambda PN}\simeq 4.5\times
10^{-12}\,{\rm GeV}^4$. Thus, at the primordial nucleosynthesis era, we
find for the above vacuum energy decay dependence, an upper limit for the
vacuum energy density is 34 orders of magnitude smaller than in previous
studies.

Due to the small values of $n_{\rm max}$ and $\nu_{max}$ our
results indicate that if the vacuum energy is decaying into CDM,
the rate of decay is extremely small.


\vskip 6mm
\noindent {\bf Acknowledgments.} R.O. thanks the Brazilian
agencies FAPESP (grant 00/06770-2) and CNPq (grant 300414/82-0)
for partial support. A.M.P thanks FAPESP for financial support
(grants 03/04516-0 and 00/06770-2).



\end{document}